\documentclass[secnumarabic, nobibnotes, nofootinbib]{ptephy_v1}
\usepackage{amsmath,amssymb}
\usepackage{graphicx}
\usepackage{mathrsfs}
\usepackage{url}
\newcommand{\bs}[1]{\ensuremath{\boldsymbol{#1}}}

\begin{document}
                                   
\title{Testing massive-field modifications of gravity via gravitational waves}

\author[1,*]{Kei~Yamada}
\affil{Department of Physics, Kyoto University, Kyoto 606-8502, Japan}

\author[1]{Tatsuya~Narikawa}

\author[1,2]{Takahiro~Tanaka}
\affil{Center for Gravitational Physics, Yukawa Institute for Theoretical Physics, Kyoto University, Kyoto 606-8502, Japan
  \email{k.yamada@tap.scphys.kyoto-u.ac.jp}}

\date{\today}

\begin{abstract}
  The direct detection of gravitational waves now provides a new channel of testing gravity theories. Despite that the parametrized post-Einsteinian framework is a powerful tool to quantitatively investigate effects of modification of gravity theory, the gravitational waveform in this framework is still extendable.  One of such extensions is to take into account the gradual activation of dipole radiation due to massive fields, which are still only very weakly constrained if their mass $m$ is greater than $10^{-16}$ eV from pulsar observations. Ground-based gravitational-wave detectors, LIGO, Virgo, and KAGRA, are sensitive to this activation in the mass range, $10^{-14}$ eV $\lesssim m \lesssim 10^{-13}$ eV. Hence, we discuss a dedicated test for dipole radiation due to a massive field using the LIGO-Virgo collaboration's open data. In addition, assuming Einstein-dilaton-Gauss-Bonnet (EdGB) type coupling, we combine the results of the analysis of the binary black hole events to obtain the 90\% confidence level constraints on the coupling parameter $\alpha_{\rm EdGB}$ as $\sqrt{\alpha_{\rm EdGB}} \lesssim 2.47$ km for any mass less than $6 \times 10^{-14}$ eV for the first time, including $\sqrt{\alpha_{\rm EdGB}} \lesssim 1.85$ km in the massless limit.
\end{abstract}

\maketitle

\section{Introduction}

The opening of gravitational wave (GW) astronomy/astrophysics allows us to approach a new test of general relativity (GR) in strong gravity regime. 
Regarding O1 and O2 by LIGO-Virgo collaboration (LVC), ten binary black hole (BH) mergers and one neutron star (NS) binary merger are summarized in the catalog GWTC-1~\cite{LIGOScientific:2018mvr}. Testing GR has been investigated by several authors using these event data and no significant deviation from GR has been reported~\cite{TheLIGOScientific:2016src, Yunes:2016jcc, Abbott:2018lct, LIGOScientific:2019fpa}. Although one of the most interesting regime to investigate is the ringdown phase~\cite{Brito:2018rfr, Nakano:2018vay}, the concrete templates of merger-ringdown phase waveform in modified gravity theories are not established yet~\cite{Berti:2018vdi}. On the other hand, the inspiral phase can be studied by using the post-Newtonian (PN) approximation.

Although a possible test is the one in the parametrized post-Einsteinian (ppE) framework, in which the so-called ppE parameters are introduced to describe modifications of gravity theory in a generic way~\cite{Yunes:2009ke, Chatziioannou:2012rf, Sampson:2013lpa, Loutrel:2014vja, Huwyler:2014gaa}, 
a waveform in this framework does not cover the whole viable extensions of gravity. One of such extensions is to consider a massive field which is coupled to a compact object through an additional charge exciting dipole radiation in the frequency range above the mass scale of the field. Such an effect can let the modification of the gravitational waveform turn on as the frequency increases during the inspiral~\cite{Sampson:2013jpa, Sampson:2014qqa}.

Spontaneous scalarization of compact objects for massless fields was first investigated by Damour and Esposito-Fare\`{e}se~\cite{Damour:1993hw}. They showed that even scalar-tensor theories which pass the present weak-field gravity tests exhibit nonperturbative strong-field deviations from GR in systems involving NSs, whereas such massless scalar fields have been strongly constrained by the recent observation of a pulsar-white dwarf binary~\cite{Antoniadis:2013pzd}. A possible extension is to add a mass term to the scalar field since the observational bounds can be avoided if the mass of the scalar field is larger than the binary orbital frequency. Such an extension to the spontaneous scalarization is discussed by several authors (e.g. see Refs.~\cite{Chen:2015zmx, Ramazanoglu:2016kul}). In the following mass ranges of the field, the coupling is constrained by several observations: (1) $m \lesssim 10^{-27}$ eV for stability on cosmological scales~\cite{Ramazanoglu:2016kul}, (2) $m \lesssim 10^{-16}$ eV from the observations of a pulsar-white dwarf binary~\cite{Antoniadis:2013pzd}, and (3) $10^{-13}$ eV $\lesssim m \lesssim 10^{-11}$ eV~\cite{Arvanitaki:2014wva, Brito:2017zvb}, which relies on the measurements of high spins of stellar mass BHs~\cite{Narayan:2013gca}.

In the presence of matter, dynamical scalarization in scalar-tensor theories has been investigated~\cite{Barausse:2012da, Palenzuela:2013hsa, Shibata:2013pra, Taniguchi:2014fqa}. In this case, dipole radiation abruptly turns on at a given threshold, where the scalar charge of a NS in a binary grows suddenly. Possible constraints on hidden sectors, such as scalar-tensor gravity or dark matter, which induce a Yukawa-type modification to the gravitational potential are discussed by several authors for future NS binary merger observations~\cite{Kopp:2018jom, Alexander:2018qzg}. However, such activation of dipole radiation in vacuum spacetimes, i.e., in the case of binary BHs has not been discussed. It may be interesting to consider Einstein-dilaton-Gauss-Bonnet (EdGB) theory with a massive scalar field, in which a BH can have a scalar hair~\cite{Yunes:2016jcc}. Very recently, Nair {\it et~al.} discussed to constrain EdGB theory from the GW observations~\cite{Nair:2019iur}. Their order-of-magnitude estimation on the constraint is $\sqrt{\alpha_{\rm EdGB}} \lesssim 1.0$ km, where $\alpha_{\rm EdGB}$ is the coupling parameter of EdGB theory, while their Fisher-estimated and Bayesian constraints are roughly $\sqrt{\alpha_{\rm EdGB}} \lesssim 6.0$ km. In addition, Tahura {\it et~al.} have independently derived constraints on a few modified theories of gravity including EdGB theory and their constraint on EdGB theory is roughly $\sqrt{\alpha_{\rm EdGB}} \lesssim 4.3$ km~\cite{Tahura:2019dgr}. Current constraint on EdGB theory is $\sqrt{\alpha_{\rm EdGB}} \lesssim 1.9$ km, which is obtained by using low-mass X-ray binaries~\cite{Yagi:2012gp}.

In this paper, we discuss activation of dipole radiation due to massive fields and modifications to gravitational waveform. We analyze the GW events in the catalog GWTC-1~\cite{GWOSCweb} to constrain the magnitude of the modification. In particular, assuming EdGB theory with a massive scalar field, we investigate the bound for the coupling parameter. Since the ground based detectors, LIGO, Virgo, and KAGRA, are sensitive at frequencies around 10-1000 Hz, gravitational-wave signals allow us to investigate the effect of the mass in the range $10^{-14}$ eV $\lesssim m \lesssim 10^{-13}$ eV, while they reduce to the massless limit for $m \lesssim 10^{-15}$ eV. We also evaluate the measurement accuracy of $\alpha_{\rm EdGB}$ by using the Fisher information matrix.

This paper is organized as follows. 
Section~\ref{Sec:ppE} briefly reviews the gravitational waveform of the inspiral phase in the ppE framework.
Section~\ref{Sec:MassiveField} presents how the gravitational waveform is modified by a massive field.
Section~\ref{Sec:Analysis} shows the results of our analysis of LVC open data of the gravitational wave catalog using the modified gravitational-wave templates and the Fisher analysis on the determination error of the coupling parameter.
Section~\ref{Sec:Summary} is devoted to summary and discussion. 
Throughout this paper we use geometric units in which $G = 1 = c$.

\section{The ppE waveform}
\label{Sec:ppE}

In this section, we briefly review the ppE waveform following Ref.~\cite{Yunes:2009ke, Tahura:2018zuq}. In the ppE framework, the waveform is formulated by considering modifications to the binding energy and GW luminosity, which correspond to conservative and dissipative corrections, respectively. Here, we employ another approach, in which we focus on the correction to the GW frequency evolution $\dot{f}$. The frequency domain (FD) ppE waveform for the inspiral phase of compact binaries is expressed as
\[
   \tilde{h} (f) =  \tilde{h}_{\rm GR} ( 1 + \alpha u^a ) e^{i \delta \Psi} ,
 \]
 where $\tilde{h}_{\rm GR}$ is the waveform in GR and its phase is given by
 \begin{align}
   \label{Eq:GRPhase}
   \Psi_{\rm GR} = 2 \pi f t_0 - \phi_0 - \frac{\pi}{4} + \frac{3}{128} u^{-5} + \cdots
 \end{align}
with the coalescence time $t_0$ and phase $\phi_0$. $\alpha u^a$ and $\delta \Psi$ represent modifications to the amplitude and the phase, respectively. We introduce
 \[
   u \equiv ( \pi \mathcal{M} f )^{1/3} ,
 \]
 where $\mathcal{M} = M \, \eta^{3/5}$ is the chirp mass with the total mass of the binary $M = m_1 + m_2$ and the symmetric mass ratio $\eta = m_1 m_2 / M^2$. We can also model the phase correction as
 \begin{align}
   \label{Eq:ppEphase}
   \delta \Psi = - \beta u^b .
 \end{align}
 $\alpha$, $\beta$, $a$, and $b$ are the ppE parameters.

 In this paper, we focus only on the dominant correction to the energy flux due to the dipole radiation, which appears in -1PN order as is known well, while the correction to the binding energy of the system is the Newtonian order~\cite{Stein:2013wza, Yunes:2016jcc}. Since only the correction to the phase accumulates through the binary evolution, we neglect the correction to the amplitude and incorporate only the correction to the evolution of the frequency $\dot{f}$. Then, let us denote $\dot{f}$ as
\begin{align}
  \dot{f} (f) &= \dot{f}_{\rm GR} (f) + \delta\!\!\:\dot{f} (f) ,
\end{align}
 where the GR term $\dot{f}_{\rm GR} (f)$ is expressed as
\[
  \dot{f}_{\rm GR} (f) = \frac{96}{5 \, \pi \mathcal{M}^2} u^{11} + \cdots .
\]
Therefore, we obtain
\begin{align}
  \label{Eq:GWphese}
  \phi (f) &= \int^{\bar{t}} 2 \pi f (t) \, d t = \int^f \frac{2 \pi f'}{\dot{f} (f')} \, d f' \notag\\
  &= \int^f \frac{2 \pi f'}{\dot{f}_{\rm GR} (f')} \left( 1 - \frac{\delta\!\!\:\dot{f}'}{\dot{f}_{\rm GR} (f')} \right) \, d f' , \\
  \label{Eq:GWtime}
  \bar{t} (f) &= \int^{\bar{t}} d t = \int^f \frac{d f'}{\dot{f} (f')} \notag\\
  &= \int^f \frac{1}{\dot{f}_{\rm GR} (f')} \left( 1 - \frac{\delta\!\!\:\dot{f} (f')}{\dot{f}_{\rm GR} (f')} \right) \, d f' ,
\end{align}
at the leading order in $\delta\!\!\:\dot{f}$. In the stationary phase approximation the gravitational waveform in the FD, $\tilde{h} (f)$, can be expressed as
\begin{align}
  \label{Eq:GWinFD}
  \tilde{h} (f) \simeq \frac{\mathscr{A} ( f )}{2 \sqrt{\dot{f}}} e^{i \left[ \phi ( \bar{t} ) - 2 \pi f \bar{t} \,\right]} ,
\end{align}
where $\mathscr{A} (f)$ is the amplitude of time domain (TD) waveform evaluated at $\bar{t} = \bar{t} (f)$. Substituting Eqs.~\eqref{Eq:GWphese} and \eqref{Eq:GWtime} into Eq.~\eqref{Eq:GWinFD}, we find
\begin{align}
  \label{Eq:ModGWofMassiveField}
  \tilde{h} (f) = \tilde{h}_{\rm GR} ( 1 + \delta \mathcal{A} ) \, e^{i \delta \Psi} 
\end{align}
at the leading order, where
\begin{align}
  \delta \mathcal{A} &= - \frac12 \frac{\delta\!\!\:\dot{f}}{\dot{f}_{\rm GR}} , \\
  \label{Eq:deltaPhi}
  \delta \Psi &= \delta \phi - 2 \pi f \delta t = 2 \pi \int^f \frac{f - f'}{\dot{f}_{\rm GR}^2 (f')} \delta\!\!\:\dot{f} (f') \, d f' .
\end{align}

\section{Massive field}
\label{Sec:MassiveField}

At the leading order in the PN expansion, we obtain the ratio between the energy losses due to the dipole radiation and the quadrupole one as (see Appendix \ref{Sec:DipoleRad})
\begin{align}
  \frac{(d E/d t)_D}{(d E/d t)_Q} = A \frac{( \omega^2 - m^2 )^{3/2}}{\omega^3} \Theta (\omega^2 - m^2) \, u^{-2} ,
  \label{Eq:FluxDipoleRad}
\end{align}
where $\omega = \pi f$ is the angular frequency of the dipole radiation, for which the dispersion relation is $\omega^2 = k^2 + m^2$ with the wave number $k$ and the mass of the field $m$, $\Theta$ is the Heaviside step function, and $A$ is a parameter which denotes the relative amplitude of the dipole radiation. In the massless limit, we have
\begin{align}
  \label{Eq:betatoA}
  \frac{(d E/d t)_D}{(d E/d t)_Q} = A u^{- 2} .
\end{align}

Since the correction to the binding energy of the system is higher order in the PN expansion,
we obtain
\begin{align}
  \label{Eq:fdot}
  \frac{\delta\!\!\:\dot{f}}{\dot{f}_{\rm GR}} \simeq \frac{(d E/d t)_D}{(d E/d t)_Q} = A \frac{( \omega^2 - m^2 )^{3/2}}{\omega^3} \Theta (\omega^2 - m^2) \, u^{-2} .
\end{align}
In the massless limit, this becomes
\begin{align}
  \label{Eq:fdotmassless}
  \frac{\delta\!\!\:\dot{f}}{\dot{f}_{\rm GR}} \simeq A \, u^{-2} .
\end{align}

\begin{figure}[htbp]
  \begin{center}
    \includegraphics[width=105mm]{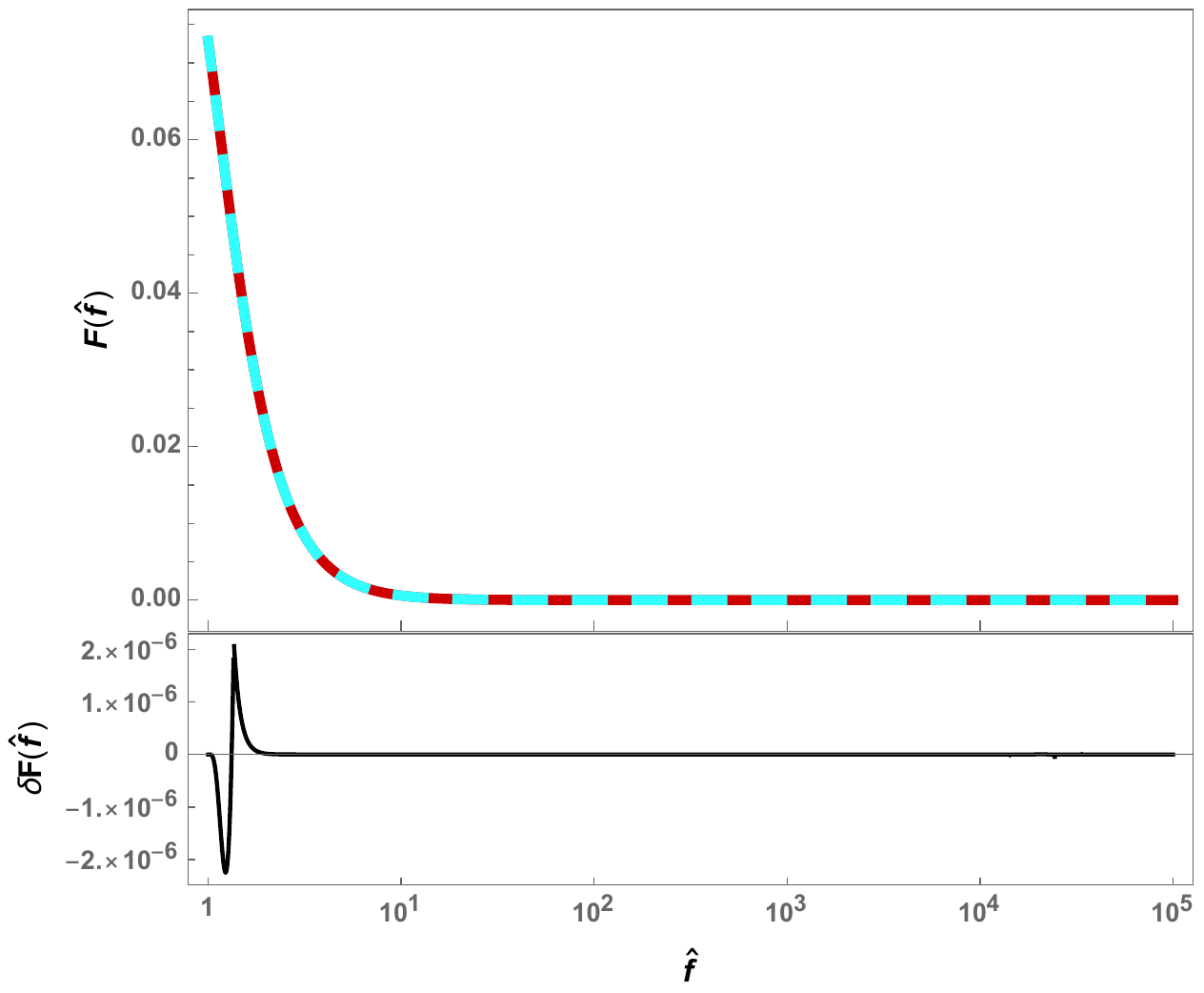}
    \caption{
      Top panel: the exact integral of Eq. \eqref{Eq:exactF} (red solid) and the fitting function, Eq.~\eqref{Eq:deltaPhiFit} (cyan dashed). Bottom: the difference of the fitting function from Eq.\eqref{Eq:exactF}. 
    }
    \label{Fig:errorfit}
  \end{center}
\end{figure}

Substituting Eq.~\eqref{Eq:fdot} into Eq.~\eqref{Eq:deltaPhi}, we obtain the modification to the GW phase as
  \begin{align}
    \label{Eq:deltaPhiOriginal}
    \delta \Psi &= 2 \pi \int_{m/\pi}^f \frac{f - f'}{\dot{f}_{\rm GR}^2 (f')} \delta\!\!\:\dot{f} (f') \, d f' \notag\\
                &\simeq \frac{5}{48} A \left( \frac{1}{m \, \mathcal{M}} \right)^{7/3} \left( F (\hat{f}) + G (\hat{f}) \right ),
  \end{align}
where we define
\begin{align}
  \label{Eq:exactF}
  F (\hat{f}) &\equiv - \int_{\hat{f}}^{\infty} d\!\!\:\hat{f}' \,  \frac{\hat{f} - \hat{f}'}{\hat{f}'^{22/3}} \left[ \hat{f}'^2 - 1 \right]^{3/2} , \\
  G (\hat{f}) &\equiv \int_1^{\infty} d\!\!\:\hat{f}' \,  \frac{\hat{f} - \hat{f}'}{\hat{f}'^{22/3}} \left[ \hat{f}'^2 - 1 \right]^{3/2} ,
\end{align}
with $\hat{f} = \pi f / m$. In our matched filtering analysis, $G (\hat{f})$ is irrelevant because
\begin{align}
G (\hat{f}) \simeq 0.0808004 \hat{f} - 0.153688 
\end{align}
can be absorbed by the shifts of the merger time and phase (see the first and the second terms in the left hand side of Eq.~\eqref{Eq:GRPhase}). Therefore, we consider only the first term $F (\hat{f})$ in the following. The asymptotic behavior of the integral, Eq.~\eqref{Eq:exactF}, becomes
  \begin{align}
    \label{Eq:deltaPhiFit}
    F (\hat{f}) \simeq
    \begin{cases}
      \vspace{0.2cm}
      \dfrac{9}{70} \hat{f}^{-7/3} \left( 1 - \dfrac{105}{208} \hat{f}^{-2} + \dfrac{105}{1672} \hat{f}^{-4} + \dfrac{1}{160} \hat{f}^{-6} + \dfrac{105}{67456} \hat{f}^{-8} \right) , &( \hat{f} \gtrsim 1.36 ) \\
      \left( \dfrac{3.12}{( \hat{f} - 1 )^{7/2}} + \dfrac{9.2}{( \hat{f} - 1 )^{5/2}} + \dfrac{8.4}{( \hat{f} - 1 )^2} + \dfrac{1}{0.08 \, ( \hat{f} - 1 )} \right)^{-1} - G (\hat{f}) . &( \hat{f} \lesssim 1.36 )
    \end{cases}
  \end{align}
In practice, we use a fitting function Eq.~\eqref{Eq:deltaPhiFit} instead of Eq.~\eqref{Eq:exactF} in order to save the computational cost. Figure~\ref{Fig:errorfit} shows the difference between Eq.~\eqref{Eq:exactF} and the fitting function \eqref{Eq:deltaPhiFit}. 

Substituting Eq.~\eqref{Eq:fdot} into Eq.~\eqref{Eq:ModGWofMassiveField}, we obtain the gravitational waveform in FD modified by the dipole radiation of a massive field, which is summarized as
\begin{align}
  \label{Eq:ModGWamp}
  \delta \mathcal{A} &= - \frac12 A \frac{( \omega^2 - m^2 )^{3/2}}{\omega^3} \, u^{-2} \, \Theta (\omega^2 - m^2) , \\
  \label{Eq:ModGWphase}
  \delta \Psi &= \frac{5}{48} A \left( \frac{1}{m \, \mathcal{M}} \right)^{7/3}  F ( \hat{f} ) \, \Theta (\omega^2 - m^2) ,
\end{align}
in Eq.~\eqref{Eq:ModGWofMassiveField} with Eq.~\eqref{Eq:deltaPhiFit}. Since we have assumed that the modification is sufficiently small, i.e. $\delta \mathcal{A} \ll 1$, and neglected the non-linear corrections, the modified gravitational waveform~\eqref{Eq:ModGWofMassiveField} becomes invalid when $\delta \mathcal{A}$ becomes large. Moreover, since we only focus on the inspiral phase in this paper, the modified waveform is not valid if the mass of the field becomes too heavy for the modifications to occur mainly in the inspiral phase. It is worth mentioning that the modified waveform, Eq.~\eqref{Eq:ModGWofMassiveField} with Eqs.~\eqref{Eq:ModGWamp} and \eqref{Eq:ModGWphase}, can model not only a scalar field modification but also the vector field one.

In this parametrization, we have two free parameters, i.e., the mass of the field $m$ and the relative amplitude of dipole radiation $A$. In general, $A$ depends on the parameters of the system such as the masses and spins of compact objects, as well as on the modified gravity theory that one considers. Applying our modified template to the actual LVC events, we will obtain the likelihood in the two parameter space $(m, A)$ for each event. To combine the results of 10 binary BH events, one should assume an extended theory which allows hairy BHs and rewrite the relative amplitude $A$ by the model parameter common to all events, such as the coupling constant of the theory. Since EdGB theory has been studied in detail and known to allow hairy BHs~\cite{Zwiebach:1985uq, Moura:2006pz, Pani:2009wy}, in this paper we consider EdGB type coupling as an example. In EdGB theory, the energy flux of the dipole radiation is expressed as~\cite{Yagi:2011xp, Berti:2018cxi, Prabhu:2018aun}
\begin{align}
  \label{Eq:betaEdGB}
    \frac{(d E/d t)_D}{(d E/d t)_Q} = \frac{5}{96} \zeta_{\rm EdGB} \frac{( m_1^2 s_2^{\rm EdGB} - m_2^2 s_1^{\rm EdGB} )^2}{M^4 \eta^{18/5}} u^{- 2} .
\end{align}
where $\zeta_{\rm EdGB} \equiv 16 \pi \alpha_{\rm EdGB}^2 / M^4$ with the coupling parameter $\alpha_{\rm EdGB}$ and
\begin{align}
    s_i^{\rm EdGB} &= \frac{2 \sqrt{1 - \chi_i^2}}{ 1 + \sqrt{1 - \chi_i^2}} & (i = 1,2)
\end{align}
is the spin-dependent factor of the BH scalar charges in the theory. Then, comparing Eqs.~\eqref{Eq:betatoA} and \eqref{Eq:betaEdGB}, we obtain the relation
\begin{align}
  \label{Eq:EdGBtypeCoupling}
  A = \frac{5 \pi}{6} \frac{\alpha_{\rm EdGB}^2}{M^4} \frac{( m_1^2 s_2^{\rm EdGB} - m_2^2 s_1^{\rm EdGB} )^2}{M^4 \eta^{18/5}} .
\end{align}

\section{Analysis}
\label{Sec:Analysis}

\subsection{Setup}
\label{Sec:Setup}

We employ the waveform Eq.~\eqref{Eq:ModGWofMassiveField} with Eqs.~\eqref{Eq:ModGWamp} and \eqref{Eq:ModGWphase} as templates to be matched with the strain data of Hanford and Livingston taken from the confident detections cataloged in GWTC-1~\cite{GWOSCweb}. Using KAGRA Algorithmic Library (KAGALI)~\cite{Oohara:2017aix}, we evaluate the likelihood, following the standard procedure of the matched filtering~\cite{maggiore2008gravitational, 2009agwd.book.....J, Creighton:2011zz} (see also Appendix \ref{Sec:MatchedFiltering}). We adopt the published noise power spectrum for each event~\cite{GWOSCweb}. The minimum and the maximum frequencies, $f_{\rm min}$ and $f_{\rm max}$, of the datasets used in the analysis are summarized in Table~\ref{Table:bstfitparameters}. As the GR waveform $\tilde{h}_{\rm GR}$, we adopt IMRPhenomD~\cite{Husa:2015iqa, Khan:2015jqa}, which is an up-to-date version of inspiral-merger-ringdown (IMR) phenomenological waveform for binary BHs with aligned spins. In this paper, since we are not interested in estimating the sky position and the inclination of the source, we choose the ratios of the $+/\times$-mode amplitudes of the respective detectors, i.e. these parameters are analytically optimized.

First, we implement a grid search to find the ``best-fit parameters'' of GR templates for each event varying the parameters $\mathcal{M}$, $\eta$, $\chi_1$, and $\chi_2$ as well as $t_0$ and $\phi_0$. The results in the detector frame are summarized in Table~\ref{Table:bstfitparameters}. Next, we calculate the likelihood for the modified templates around the GR best-fit. As is well known, there is an approximate degeneracy among the mass ratio and the spins in the inspiral phase waveform. In fact, we find that it is unnecessary to take into account the both variations in our calculations. Therefore, here we fix the spins to save the computational costs (see also Appendix \ref{Sec:DegeneracySpin}).

Moreover, to quantify the measurement accuracy of $\alpha_{\rm EdGB}$, we compute the Fisher information matrix defined by
\begin{eqnarray}
 \Gamma_{i j} \equiv \left( \left. \frac{\partial \hat{h}}{\partial \theta^i} \right| \frac{\partial \hat{h}}{\partial \theta^j} \right) ,
\nonumber
\end{eqnarray}
where $\hat{h}$ is an appropriately normalized template and $\theta^i$ are its parameters. For a large SNR, the root-mean-square error in $\theta^i$ can be evaluated as
\begin{eqnarray}
 \Delta \theta^i \equiv \sqrt{\langle ( \theta^i )^2 \rangle} = \sqrt{( \Gamma^{- 1} )^{i i}}.
\nonumber
\end{eqnarray}
We use a polynomial fit given in Eq.~(C.1) in Ref.~\cite{Yunes:2016jcc} as the noise power spectrum density of Advanced LIGO around GW151226 and GW170608. We take the lowest frequency to be $f_{\rm low}=$10 Hz. We use the restricted TaylorF2 PN aligned-spin model as the waveform model \cite{Buonanno:2009zt, Arun:2008kb, Mikoczi:2005dn, Khan:2015jqa}.

\begin{table}[htb]
  \begin{center}
    \caption{The GR ``best-fit parameters'' in the detector frame}
  \begin{tabular}{c c c c c c c} \hline
    event & $(f_{\rm min}, f_{\rm max})$/Hz & $\mathcal{M}/M_{\odot}$ & $\eta$ & $\chi_1$ & $\chi_2$ & SNR \\
    \hline
    \hline
    GW150914 & (20, 1024) & 31.2 & 0.249 &  0.79 & -1.00 & 24.4 \\
    GW151012 & (20, 1024) & 18.0 & 0.249 & -0.36 &  0.22 & 9.00 \\
    GW151226 & (20, 1024) & 9.70 & 0.211 &  0.50 & -0.46 & 12.0 \\
    GW170104 & (20, 1024) & 24.9 & 0.243 & -0.87 &  1.00 & 13.2 \\
    GW170608 & (20, 1024) & 8.48 & 0.250 &  0.47 & -0.49 & 15.4 \\
    GW170729 & (20, 1024) & 48.4 & 0.201 &  0.74 & -1.00 & 10.4 \\
    GW170809 & (20, 1024) & 30.5 & 0.228 &  0.72 & -1.00 & 12.0 \\
    GW170814 & (20, 1024) & 26.9 & 0.246 &  0.85 & -1.00 & 16.2 \\
    GW170817 & (23, 2048) & 1.20 & 0.231 & -0.41 &  0.96 & 32.2 \\
    GW170818 & (16, 1024) & 32.9 & 0.250 &  0.98 & -1.00 & 10.5 \\
    GW170823 & (20, 1024) & 39.3 & 0.249 &  0.99 & -1.00 & 11.5 \\
    \hline
  \end{tabular}
    \label{Table:bstfitparameters}
  \end{center}
\end{table}

\subsection{Results}

\begin{figure}[htbp]
  \begin{center}
    \includegraphics[width=105mm]{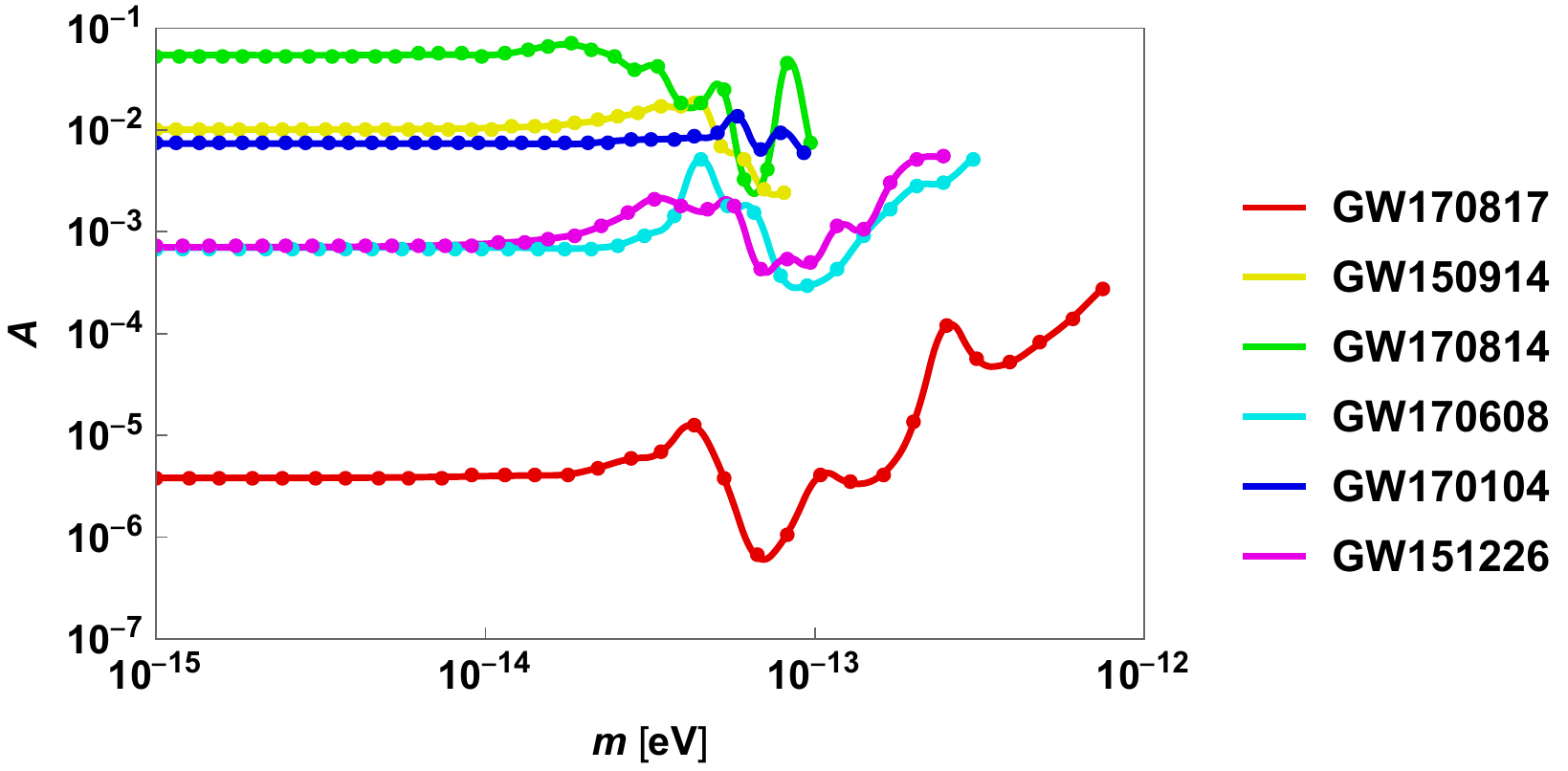}
    \caption{The 90\% CL constraints on $A$ for six events possessing relatively high SNR. The calculations are terminated when the mass scale reaches a threshold frequency of each event, which is chosen to be ${\rm min}(0.5 f_{\rm peak}, f_{\rm max})$}
    \label{Fig:90CLA}
  \end{center}
\end{figure}

Figure~\ref{Fig:90CLA} shows the 90\% confidence level (CL) constraints on $A$ for each event. To obtain the constraints, we calculate the likelihood, Eq.~\eqref{Eq:likelihood}, of the best fit template for each point of $(A,m)$ plane. The likelihood can be regarded as the unnormalized posterior distribution when the prior of $A$ is uniform as well as those of $\mathcal{M}$ and $\eta$. Thus, we integrate it with respect to $A$ for each $m$ to evaluate 90\% CL constraints. Here, we show only six events possessing relatively high SNR because the constraints on $A$ obtained from the other events are much larger and they are in the region where the error due to linearization becomes too large. Since our modified template becomes no longer valid if the mass of the field is too heavy, we terminate the calculations when the mass scale reaches a threshold frequency of each event, which is chosen to be ${\rm min}(0.5 f_{\rm peak}, f_{\rm max})$, where $f_{\rm peak}$ is the peak frequency of GWs. Because of the existence of the noise, the constraints seem to oscillate at the frequency corresponding to the region $m \gtrsim 5 \times 10^{-14}$ eV, where the detectors are sensitive to the mass effects. For all events, larger values of $A$ remain unconstrained for heavier masses because the frequency that the dipole radiation turns on is too high for the detectors to be sensitive. This figure indicates that the constraints tend to be stronger for lighter chirp mass events. This is reasonable because events with a lighter chirp mass have longer inspiral phase at higher frequencies and the dipole radiation is basically -1PN correction which becomes more efficient in the case of a lighter chirp mass for fixed $A$ and $f$ as one can see in Eq.~\eqref{Eq:FluxDipoleRad}. Our constraint on $A$ for GW170817 is consistent with the estimated constraint in Ref.~\cite{Alexander:2018qzg}\footnote{In their paper, the magnitude of modification due to the dipole radiation is governed by $\gamma$, which is related with $A$ as \[ \gamma = \frac{48}{5 \eta^{2/5}} A. \] For a binary NSs with SNR = 25 for the design sensitivity of LIGO, they estimated the constraint as $\gamma \lesssim 10^{-4}$ , which corresponds to $A \lesssim 10^{-6}$.}. We should notice that in general a strong constraint on $A$ indicates that the modification to the waveform is strongly constrained, but it does not always imply a strong constraint on the modification to theory. This is because in general in scalar-tensor theory $A$ is proportional to the squared difference of the scalar charges of constituents of the binary as shown in Eq.~\eqref{Eq:EdGBtypeCoupling}. Therefore, if the difference vanishes, the modification to the waveform vanishes, i.e., $A = 0$, even for a quite large coupling constant, and no one can constrain the modification to such a theory.

\begin{figure}[htbp]
  \begin{center}
    \includegraphics[width=105mm]{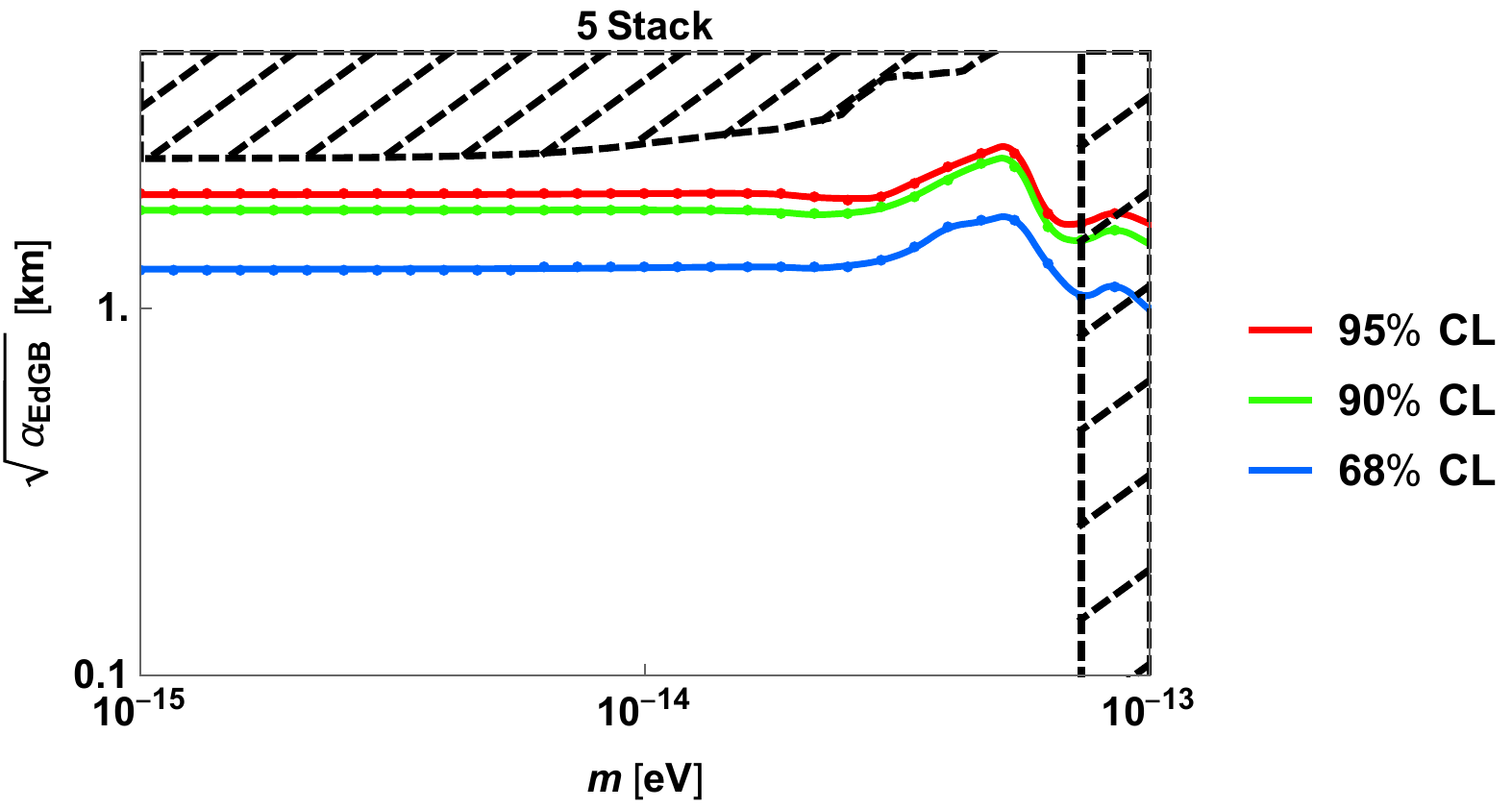}
    \caption{
      The 68\% CL, 90\% CL, 95\% CL constraints on $\sqrt{\alpha_{\rm EdGB}}$ after combining five binary BH events possesing relatively high SNR, i.e. GW150914, GW170814, GW170608, GW170104, GW151226, by assuming the EdGB type coupling~\eqref{Eq:EdGBtypeCoupling}. 
    }
    \label{Fig:90CLalpha}
  \end{center}
\end{figure}

By combining the results, we can obtain more stringent and reliable constraints. Figure~\ref{Fig:90CLalpha} shows the 68\% CL, 90\% CL, 95\% CL constraints on $\sqrt{\alpha_{\rm EdGB}}$ after combining five binary BH events possesing relatively high SNR, i.e. GW150914, GW170814, GW170608, GW170104, GW151226, by assuming the EdGB type coupling~\eqref{Eq:EdGBtypeCoupling}. Similar to Fig.~\ref{Fig:90CLA}, assuming the prior distribution of $\sqrt{\alpha_{\rm EdGB}}$ is uniform, we derive the confidence intervals for $\sqrt{\alpha_{\rm EdGB}}$ for each $m$ from the integral of the likelihood. We plot the regions $10^{-15}$ eV $\lesssim m \lesssim$ $10^{-13}$ eV and $0.1$ km $\leq \sqrt{\alpha_{\rm EdGB}} \leq$ $5$ km. In the case of EdGB type of coupling the dipole modification vanishes if $m_1^2 s_2^{\rm EdGB} = m_2^2 s_1^{\rm EdGB}$. Therefore, in this case the leading signature of modification vanishes. In our calculations, we find no event which satisfies $m_1^2 s_2^{\rm EdGB} = m_2^2 s_1^{\rm EdGB}$. However, recalling that spins are almost zero-consistent, i.e., $s_i \simeq 1$, one can find it is very difficult to constrain the modification from nearly-equal mass binaries, such as GW170818. The regions where error due to nonlinearity, $\left( \delta \mathcal{A} \right)^2$, becomes large [$\left( \delta \mathcal{A} \right)^2 \sim 0.01$] and the frequency at which the modification occurs exceeds $0.5 f_{\rm peak}$ are hatched by black dashed lines. Here, we stacked only five events. This is because the region where our approximations are valid becomes smaller by including heavier chirp mass events, such as GW170729, which do not improve the constraints. Strictly speaking, the parameters in the hatched region are not excluded. However, the likelihood rapidly decreases before the hatched region and it seems quite unnatural that the likelihood increases in the hatched region to give the maximum even if we use templates which are valid there. Therefore, we believe that our constraints here are appropriate. We find that $\sqrt{\alpha_{\rm EdGB}}$ is constrained as $\sqrt{\alpha_{\rm EdGB}} \lesssim 2.47$ km for all mass below $6 \times 10^{-14}$ eV for the first time with 90\% CL including $\sqrt{\alpha_{\rm EdGB}} \lesssim 1.85$ km in the massless limit. This constraint in the massless limit is much stronger than the results of in Ref.~\cite{Nair:2019iur, Tahura:2019dgr}, while it is accidentally almost the same with that by low-mass X-ray binaries~\cite{Yagi:2012gp}.

In order to investigate this disagreement with Ref.~\cite{Nair:2019iur, Tahura:2019dgr}, we calculated the Fisher matrix with respect to the source parameters $\{ \ln d_L, t_0, \phi_0, {\cal M}, \eta, \chi_1, \chi_2, \alpha_{\rm EdGB}^2\}$ where $d_L$ is the luminosity distance. In Table \ref{table:Fisher}, we show the estimated constraints on $\alpha_{\rm EdGB}^2$, the measurement accuracy of $\eta$, and the correlation coefficient between $\eta$ and $\alpha_{\rm EdGB}^2$ in the cases of GW151226-like and GW170608-like signals, which are modeled by the inspiral PN waveform TaylorF2. The Fisher matrix is evaluated at the median value of the posterior distributions of GWTC-1 catalog~\cite{LIGOScientific:2018mvr}. The lower layer in Table \ref{table:Fisher} shows the case when all parameters are unconstrained a priori, while the upper layer takes into account the prior information that $\chi_1$ and $\chi_2$ must necessarily be smaller than unity, following Ref.~\cite{Cutler:1994ys} (see also Ref.~\cite{Poisson:1995ef}). The presented values are for a single detector. When we use spin prior, constraints on $\sqrt{\alpha_{\rm EdGB}}$ are about 40\% stronger than that when we do not, for both GW151226-like and GW170608-like signals.

The constraints without spin prior are consistent with those presented previously in Ref.~\cite{Nair:2019iur, Tahura:2019dgr}, while the constraints with spin prior are in good agreement with our main analysis with strain data. GW170608-like signal gives stronger constraints on $\sqrt{\alpha_{\rm EdGB}}$ than GW151226-like signal. This is because the measurement accuracy of $\eta$ for GW151226-like signal is worse than that for GW170608-like signal, and the correlation between the symmetric mass-ratio and $\sqrt{\alpha_{\rm EdGB}}$ for GW151226-like signal is larger than that of GW170608-like signal. Finally, we should note that while the results of Fisher analysis imply that incorporating the spin prior affects the constraints by a factor, but the degeneracy between the spins and $\alpha_{\rm EdGB}$ is not so important as long as variations of the spins are limited in a relevant range as shown in Fig.~\ref{Fig:90CLspin}.

\begin{table}[htb]
  \begin{center}
    \caption{
      Fisher-estimated constraints on EdGB gravity model from GW151226-like and GW170608-like signals. The lower layer shows the case when all parameters are unconstrained, while the upper layer takes into account the prior information that $\chi_1$ and $\chi_2$ must necessarily be smaller than unity. The measurement accuracy of symmetric mass-ratio and correlation coefficient between $\eta$ and $\sqrt{\alpha_{\rm EdGB}}$ are also shown.
    }
    \begin{tabular}{c c c c} \hline
      signal & $\sqrt{\alpha_{\rm EdGB}}$ [km] & $\Delta\eta/\eta$ & $c_{\eta,\sqrt{\alpha_{\rm EdGB}}}$  \\
      \hline \hline
      \multicolumn{4}{c}{with spin prior} \vspace*{2pt} \\
      GW151226-like & 3.35 & 361\% & -0.708 \\
      GW170608-like & 2.96 & 120\% & -0.245\\
      \hline
      \multicolumn{4}{c}{without spin prior} \vspace*{2pt} \\
      GW151226-like & 5.80 & 11600\% & 0.968 \\
      GW170608-like & 5.23 & 7110\% & 0.957 \\
      \hline
    \end{tabular}
    \label{table:Fisher}
  \end{center}
\end{table}

\section{Summary and discussion}
\label{Sec:Summary}

\begin{figure}[htbp]
  \begin{center}
    \includegraphics[width=105mm]{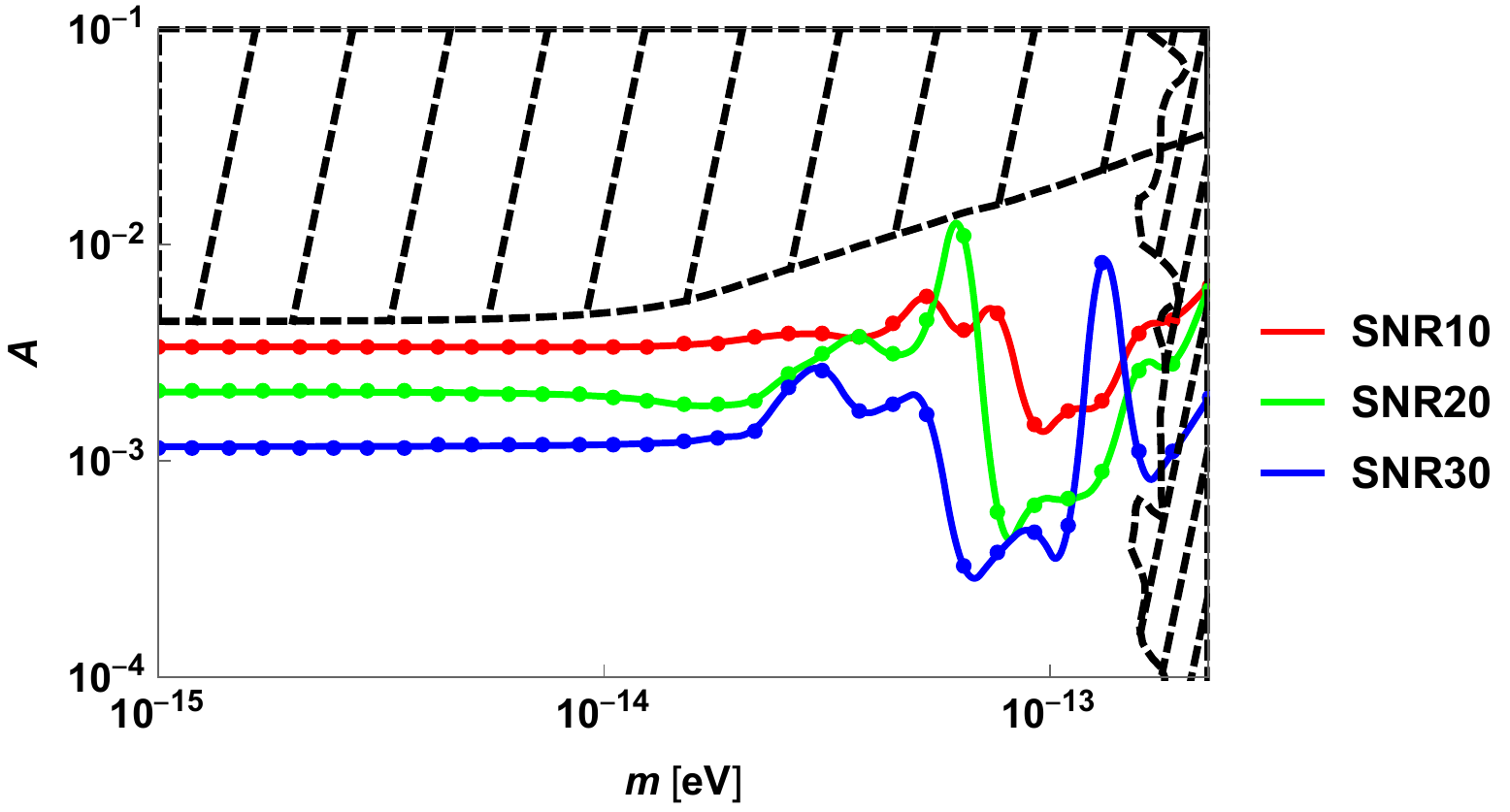}
    \caption{
      The 90\% CL constraints on $A$ for each mock data. This implies 90\% CL constraints are approximately in inverse proportion to SNR.
      }
    \label{Fig:90CLmock}
  \end{center}
\end{figure}

\begin{figure}[htbp]
  \begin{center}
    \includegraphics[width=150mm]{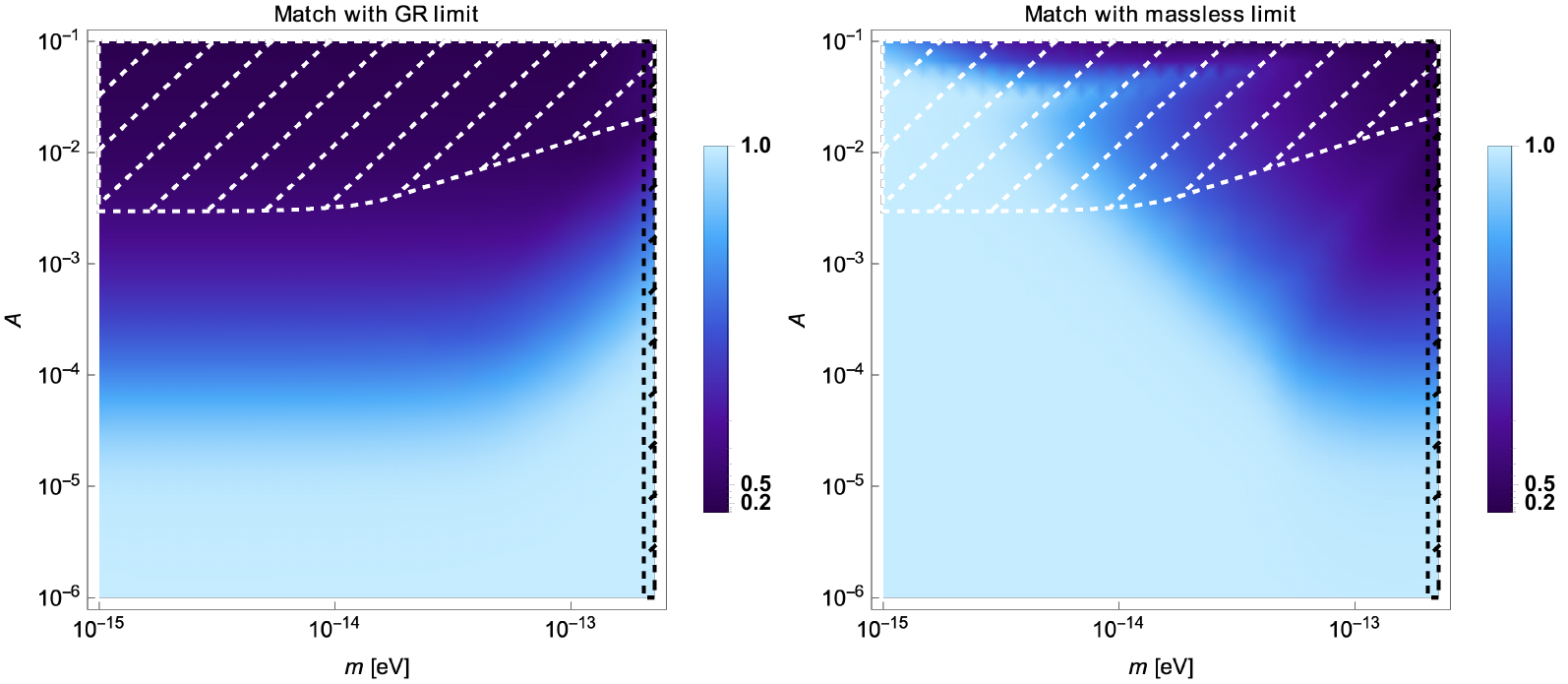}
    \caption{
      Left panel: match between the modified template due to a massive field and the GR limit. Right panel: the modified template due to a massive field and the massless limit.
    }
    \label{Fig:matchmock}
  \end{center}
\end{figure}

In order to investigate the role of the noise in our calculations in the parameter space $(m, A)$, we prepare three mock datasets in which an artificial GR IMRPhenomD waveform is injected into the Gaussian noise with SNR about 10, 20, and 30, respectively. The parameters of the waveform are fixed to
$\mathcal{M} = 10.6 \, M_{\odot}$ in the detector frame, $\eta = 0.24$ (so that $m_1 = 15 \, M_{\odot}$ and $m_2 = 10 \, M_{\odot}$), and $\chi_1 = \chi_2 = 0$. Figure~\ref{Fig:90CLmock} shows the 90\% CL constraints on $A$ for each mock data. Here, assuming the uniform prior distribution of $A$, we integrate the likelihood with respect to $A$ for each $m$ as we did in Figs.~\ref{Fig:90CLA} and \ref{Fig:90CLalpha}. The regions where our approximation breaks down are hatched by black dashed lines. This indicates 90\% CL constraints are approximately in inverse proportion to SNR for lighter $m$. Similarly to Fig. \ref{Fig:90CLA}, the constraints seem to oscillate at the frequency corresponding to the region $m \gtrsim 5 \times 10^{-14}$ eV because of the existence of the noise. The 90\% CL boundaries for different SNR cases cross with each other in the range with heavier mass of the field, because we use a different noise realization in each case. If we use the same noise but with different magnitudes of signals, then the constraints are just scaled. As mentioned above, it seems quite unnatural that the likelihood increases in the hatched region to give the maximum even if we use templates which are valid there, although logically this possibility cannot be completely excluded. Therefore, 90\% CL constraints would be still appropriate even if the constraints are close to the border of validity region (see also Appendix \ref{Sec:Injection}). Since we use only one realization of injection for mock data, the result may depend on the noise realization. To investigate statistical feature by examining a number of realizations is left for a future work.

Figure~\ref{Fig:matchmock} shows the matches [Eq.~\eqref{Eq:match}] between the template modified by the effect of a massive field and the GR limit (left panel) and between the former and the massless limit for each value of $A$ (right panel), where we use the same values for $\mathcal{M}$, $\eta$, $\chi_1$ and $\chi_2$. The regions where our approximation is invalid are hatched by white dashed and black dashed lines, respectively. From the match with the GR limit, larger values of $A$ tend to be allowed for heavier masses even if there is no deviation from GR because the threshold frequency beyond which the dipole radiation is excited is too high and the detectors are less sensitive there. This is consistent with Fig.~\ref{Fig:90CLA}. On the other hand, the right panel shows that as expected it becomes difficult to distinguish a massive field from a massless field if the mass of the field gets smaller, while the match gets drastically smaller if the field is massive enough. Therefore, we cannot substitute the massless templates for the massive one if we try to constrain the magnitude of modification starting with the frequency band of the ground base detectors.

We discussed a dedicated test of the massive-field modification using the LIGO/Virgo's open data of GWTC-1. In addition, assuming EdGB type coupling, we stacked the results of five events possessing relatively high SNR, and found the 90\% CL constraints on the coupling parameter $\alpha_{\rm EdGB}$ as $\sqrt{\alpha_{\rm EdGB}} \lesssim 2.47$ km for any mass less than $6 \times 10^{-14}$ eV for the first time including $\sqrt{\alpha_{\rm EdGB}} \lesssim 1.85$ km in the massless limit. In this analysis, we choose the ratios of the $+/\times$-mode amplitudes of the respective detectors, i.e. these parameters are analytically optimized. Moreover, since the number of events with a lighter chirp mass, such as binary NS, must increase in the near future, it is interesting to stack those events by assuming various theories in which charged NSs are allowed consistently. Furthermore, the NS-BH binary is also interesting~\cite{Carson:2019fxr}. Because, in addition to their light chirp mass, $A$ can be large in the theory that only either NS or BH has a charge, for example EdGB theory. Moreover, future multiband observations will improve the current upper bounds of various modified theories~\cite{Gnocchi:2019jzp}.

\section*{Acknowledgments}
The authors are grateful to the developers of KAGRA Algorithmic Library (KAGALI), especially to Hideyuki~Tagoshi, Nami~Uchikata, and Hirotaka~Yuzurihara for their support. The authors would like to thank Hiroyuki~Nakano for the useful conversations and comments. K.Y. thanks to Nicol\'{a}s~Yunes and his group for their hospitality and nice conversations. K.Y. also wish to acknowledge the Yukawa Institute for Theoretical Physics at Kyoto University, where this work was developed during ``The 3rd Workshop on Gravity and Cosmology by Young Researchers'' (YITP-W-18-15). This work was supported by JSPS KAKENHI Grant Number JP17H06358 (and also JP17H06357), A01: {\it Testing gravity theories using gravitational waves}, as a part of the innovative research area, ``Gravitational wave physics and astronomy: Genesis''. T.N.'s work was also supported in part by a Grant-in-Aid for JSPS Research Fellows. T. T. acknowledges support from JSPS KAKENHI Grant No. JP15H02087. 

\appendix
\section{Dipole Radiation}
\label{Sec:DipoleRad}

By separation of variables $\phi (t, \bs{r}) = \varphi (t) \psi (\bs{r})$, the scalar field equation is decomposed to a set of equations as
\begin{align}
  \left( \frac{d}{d t^2} + \omega^2 \right) \varphi (t) &= 0 , \\
  \left( \nabla^2 + k^2 \right) \psi (\bs{r}) &= S (\bs{r} ) ,
  \label{SecApp1Eq:Helm}
\end{align}
where $\omega^2 = k^2 + m^2$. Here we assume the validity of the PN expansion and hence the source $S (\bs{r} )$ is a function of $\bs{r}$. The Green's function of Eq.~\eqref{SecApp1Eq:Helm} is expanded by spherical harmonics as
\begin{align}
  \frac{e^{i k |\bs{r} - \bs{r'}|}}{4 \pi |\bs{r} - \bs{r'}|} = i k \sum_{\ell = 0}^{\infty} j_{\ell} (k r') h_{\ell}^{(1)} (k r) \sum_{m = - \ell}^{\ell} Y_{\ell m}^*(\Omega') Y_{\ell m}(\Omega) ,
\end{align}
where $r = |\bs{r}|$, $r' = |\bs{r}'|$, we suppose $r > r'$, and $j_{\ell}$ and $h^{(1)}_{\ell}$ are, respectively, the spherical Bessel function and the spherical Hankel function. Therefore, the solution of Eq.~\eqref{SecApp1Eq:Helm} can be expressed as
\begin{align}
  \phi(t, \bs{r}) \sim i k \sum_{\ell = 0}^{\infty} \sum_{m = - \ell}^{\ell} \int d^3 x' \, e^{- i \omega (t - t')} S( \bs{r}' ) (k r')^{\ell} h_{\ell}^{(1)} (k r) Y_{\ell m}^*(\Omega') Y_{\ell m}(\Omega) .
\end{align}

Thus, the $k$-dependence of the energy loss by the dipole radiation of the scalar can schematically be written as
\begin{align}
  \label{Eq:kdependencedEdt}
  \frac{d E}{d t} &= \int T^{\phi}_{t r} \, d \Sigma \propto \frac{\partial \phi}{\partial t} \, \frac{\partial \phi}{\partial r} \propto k^4 \frac{\partial \left( h_1^{(1)} (k r) \right)^2}{\partial r} \propto k^3  &\text{for $\ell = 1$} ,
\end{align}
where $T^{\phi}_{\mu \nu}$ is the energy-momentum tensor of the scalar field, $d \Sigma$ is the surface element, and we assume $k r' \ll 1$ and expand the spherical Bessel function. The dot and dash denote the derivatives with respect to time and radial coordinates, respectively. The order of magnitude of dipole radiation is larger than that of quadruple because the energy flux of the dipole and quadrupole radiation can be estimated as
\begin{align}
  \frac{(d E/d t)_D}{(d E/d t)_Q} &\sim \frac{\langle \ddot{D} \ddot{D} \rangle}{\langle \dddot{Q} \dddot{Q} \rangle} \sim \frac{\left( r \omega^2 \right)^2}{\left( r^2 \omega^3 \right)^2}  \propto u^{-2} ,
  \label{Eq:quaddEdt}
\end{align}
where $D$ and $Q$ are the dipole and quadruple moments, respectively, $\langle \cdots \rangle$ denotes a temporal average over several periods of GWs, and we have used $\omega = \omega_{\rm orbit} \propto r^{-3/2}$ with the orbital angular velocity $\omega_{\rm orbit}$. Combining Eqs.~\eqref{Eq:kdependencedEdt}--\eqref{Eq:quaddEdt}, the energy loss by dipole radiation can be estimated as
\begin{align}
  \frac{(d E/d t)_D}{(d E/d t)_Q} = A \frac{k^3}{\omega^3} \Theta (\omega^2 - m^2) \, u^{-2} ,
\end{align}
where $A$ is an overall factor and we introduce the step function to represent the sudden activation of dipole radiation.

\section{Matched Filtering}
\label{Sec:MatchedFiltering}

Define the scalar product between two real functions $h(t)$ and $g(t)$ by
\begin{align}
  ( h|g ) \equiv 4 \operatorname{Re} \int_0^{\infty} d f \frac{\tilde{h} (f) \, \tilde{g}^* (f)}{S_n (f)} ,
  \label{Eq:match}
\end{align}
where $\operatorname{Re}$ denotes the real part, $\tilde{}$ indicates the Fourier transformation of the TD functions, and $S_n (f)$ is the noise power spectrum density as
\[
  \langle \tilde{n}(f) \tilde{n}^*(f') \rangle = \delta(f - f') \frac12 S_n (f) .
\]
The Gaussian probability distribution for the noise $\tilde{n}$ is
\[
  p(\tilde{n}) \propto \exp \left[ -\frac12 ( n|n ) \right] .
\]
We are assuming that the output of the detector satisfies the condition for claiming detection, i.e. it is of the form $s(t) = \rho h(t; \theta) + n_0(t)$, where $n_0$ is a specific realization of the noise, and $h(t;\theta)$ is the signal with the parameter $\theta$ and the SNR of the event $\rho = ( s|h )$ is sufficiently high. The likelihood function for the observed output $s(t)$, given that there is a GW signal corresponding to the parameters $\theta$, is obtained plugging $n_0 = s - \rho h(\theta)$ into the above equation,
\begin{align}
  \Lambda(s|\theta) \propto \exp \left[ -\frac12 ( s - \rho h(\theta) | s - \rho h(\theta) ) \right]
  \propto \exp \left[ \frac12 \rho^2 \right] .
  \label{Eq:likelihood}
\end{align}

\section{Approximate degeneracy among the mass ratio and the spins}
\label{Sec:DegeneracySpin}

\begin{figure}[htbp]
  \begin{center}
    \includegraphics[width=120mm]{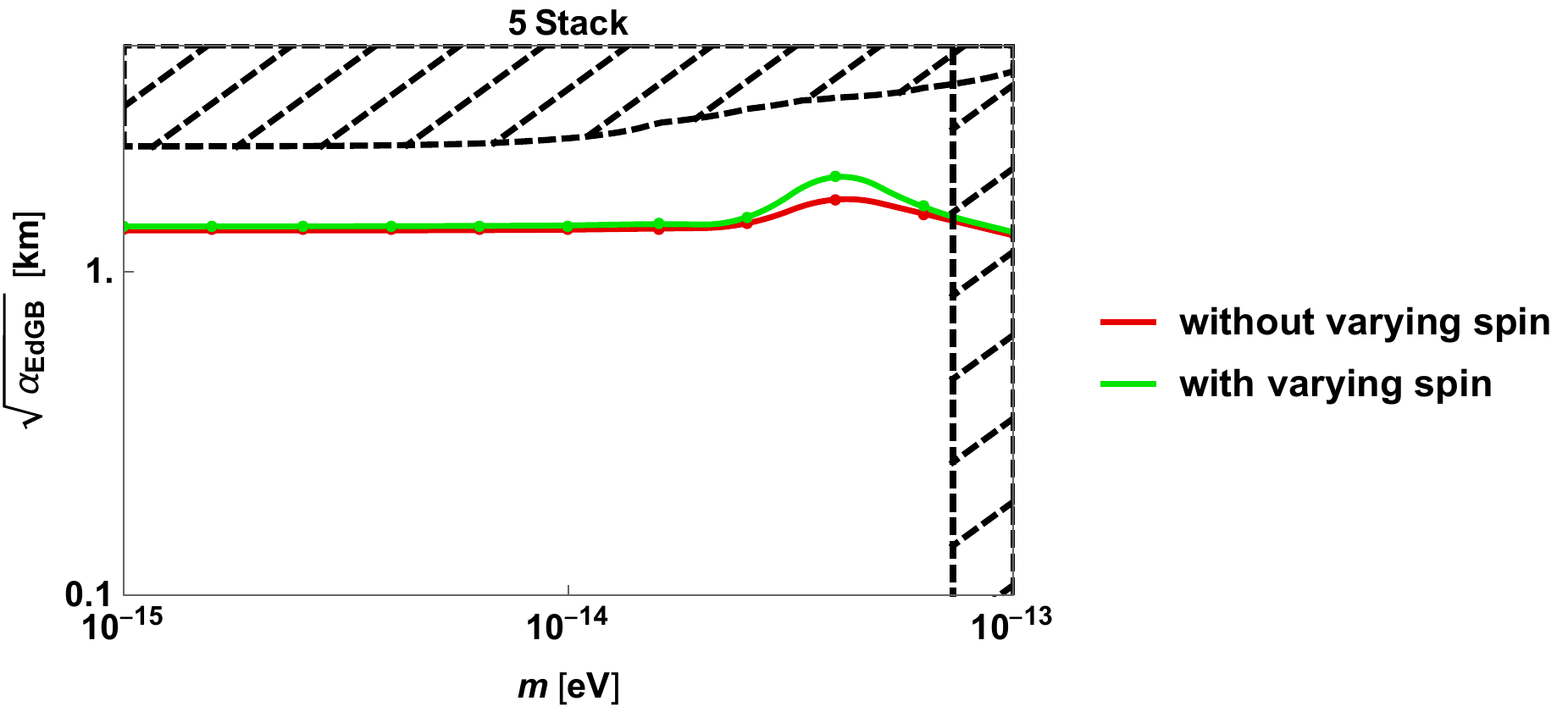}
    \caption{Comparison of the 90\% CL constraints on $\sqrt{\alpha_{\rm EdGB}}$ between Case 1 and Case 2, in which we implement a grid search with/without varying the spins, respectively.
    }
    \label{Fig:90CLspin}
  \end{center}
\end{figure}

To investigate the effect of the spins on the constraints, we compare the 90\% CL constraints on $\sqrt{\alpha_{\rm EdGB}}$ in two cases, after combining five binary BH events possesing relatively high SNR. Case 1: the same with the 90\% CL constraint in Fig.~\ref{Fig:90CLalpha}. Case 2: we additionary vary the spins and estimate the 90\% CL constraint. In Case 2, we reduce the number of sample points to save the computational costs. Figure~\ref{Fig:90CLspin} shows the comparison between them. It is found that the effect of the spins on the constraints is small enough. 

\section{Effects on SNR by increasing the number of parameters}
\label{Sec:NofParam}

How much SNR can become larger if the number of parameters increases? To see this, let us consider the number of degrees of freedom in the parameter search. Let $(x_i|x_j)$ denote the match between the datasets $x_i$ and $x_j$. Substituting $x_i = s - \rho_i \hat{h}_i = \rho \hat{h} - \rho_i \hat{h}_i + n$, we have a relation between the match and the chi-squared as 
\begin{align}
  \chi_i^2
  &= (x_i|x_i) 
  = ( s | s ) - | \rho_i |^2 ,
\end{align}
where we assume $\hat{h}_i$ is chosen to be the best fit within a given template bank. Therefore, the difference of $\chi^2$ between different template banks is
\begin{align}
  \chi_1^2 - \chi_2^2 = | \rho_2 |^2 - | \rho_1 |^2 .
\end{align}
Let $i = 1, 2$ represent the GR and modified theory, respectively. In general, increasing the number of parameters causes decreasing the number of degrees of freedom of $\chi^2$-distribution, so that the value $\chi^2$ decreases. Therefore, the mean value of the left-hand side of the above equation becomes 2. On the other hand, the right-hand side of the above equation is the deference of SNR. For example, we obtain $| \rho_2 |^2 - | \rho_1 |^2 = 2.13934$ by averaging all eleven GW events.

\section{Injection test of detectability of massive field modification}
\label{Sec:Injection}

\begin{figure}[htbp]
  \begin{center}
    \includegraphics[width=150mm]{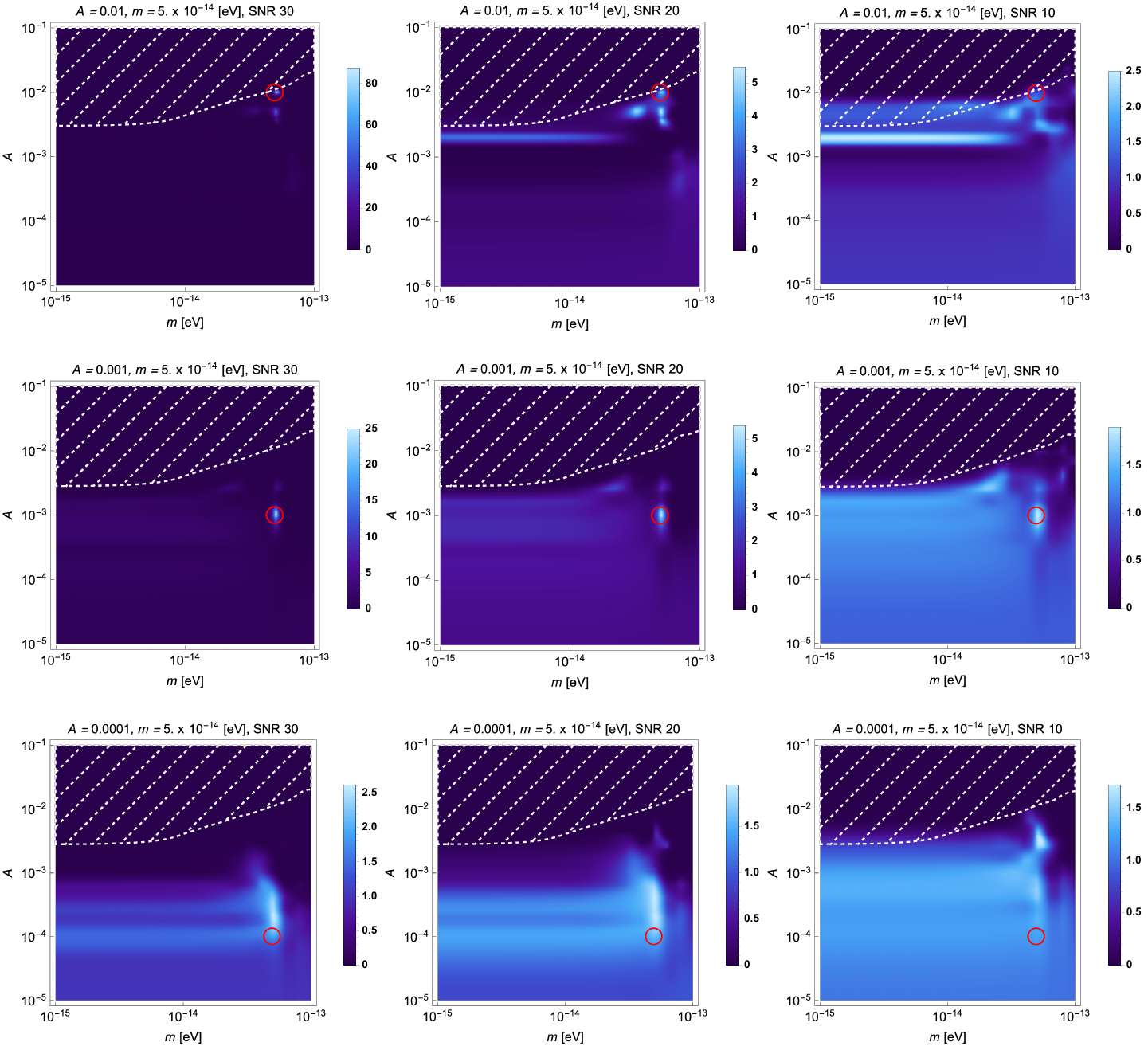}
    \caption{The likelihood normalized by the GR values in the $(A, m)$ plane for each mock data. The red circle shows the injected values of $A$ and $m$. This figure implies that we can detect the modification due to the massive field if $A > 10^{-3}$ and SNR is larger than 20, while it is possible to constrain $A$ for the other cases.}
    \label{Fig:InjectionModifiedTemp}
  \end{center}
\end{figure}

In order to demonstrate the detectability of the modification caused by a massive field, we prepare nine mock datasets for which artificial modified waveforms are injected into the Gaussian noise with $m = 5 \times 10^{-14}$ eV, $A = 10^{-2}, 10^{-3}$, and $10^{-4}$ and SNR about 10, 20, and 30. The other parameters of the waveform are fixed to $\mathcal{M} = 10.6 \, M_{\odot}$ in the detector frame, $\eta = 0.24$ (so that $m_1 = 15 \, M_{\odot}$ and $m_2 = 10 \, M_{\odot}$), and $\chi_1 = \chi_2 = 0$. Figure~\ref{Fig:InjectionModifiedTemp} shows the likelihood normalized by the GR values in the $(A, m)$ plane for each mock data. The region hatched by white dashed lines indicates the outside of the range of validity. The red circle shows the injected values of $A$ and $m$. In general, as we discussed in Appendix~\ref{Sec:NofParam}, increasing the number of parameters of templates leads to increasing SNR. In our case, since the number of the additional parameters is two, the likelihood normalized by the GR values can reach to $e \simeq 2.7$ regardless of whether the signal is GR or not. Therefore, the likelihood needs to be much larger than this value for a confident detection of the modification. From this point of view, Fig.~\ref{Fig:InjectionModifiedTemp} implies that for the case of $\mathcal{M} = 10.6 \, M_{\odot}$ we may detect the modification due to the massive field if $A \gtrsim 10^{-3}$ and SNR is larger than 20. For the other cases it is possible only to give an upper bound to $A$. However,in the case that $A = 10^{-2}$ with SNR is 10 (Right-Top panel in Fig.~\ref{Fig:InjectionModifiedTemp}), the likelihood remains large even at the boundary of the hatched region and hence a meaningful constraint will not be obtained.
Since we use only one realization of injection for mock data, the result may depend on the noise realization. To investigate statistical feature by a number of realizations is left for a future work.

\bibliographystyle{ptephy}
\bibliography{Ref}

\end{document}